\documentclass[preprintnumbers,amsmath,amssymb]{revtex4}

\usepackage{graphicx}
\usepackage{times}
\usepackage{isolatin1}

\def\ie{{\it i.e.\;}}
\def\eg{{\it e.g.\;}}
\def\GeV{{\;\rm GeV}}

\def\msbar{$\overline{\rm MS}$ }
\def\hh{H^+H^-}
\def\bb{\bar{b}}

\begin{document}

\preprint{$
\begin{array}{l}
\mbox{hep-ph/0503135}\\
\mbox{MPP-2005-15}
\end{array}
$}

\title{Charged Higgs Boson Pairs at the LHC}

\author{Alexandre Alves$^1$ and 
        Tilman Plehn$^2\footnote{Heisenberg Fellow.}$}
\affiliation{$^1$Instituto de Física, Universidade de São Paulo, São Paulo, Brasil}
\affiliation{$^2$Max Planck Institute for Physics, Munich, Germany}\bigskip


\begin{abstract} 
 We compute the cross section for pair production of charged Higgs
 bosons at the LHC and compare the three production mechanisms. The
 bottom--parton scattering process is computed to NLO, and the
 validity of the bottom--parton approach is established in detail. The
 light--flavor Drell--Yan cross section is evaluated at NLO as
 well. The gluon fusion process through a one--loop amplitude is then
 compared with these two results. We show how a complete sample of
 events could look, in terms of total cross sections and
 distributions of the heavy final states.
\end{abstract}


\maketitle

\section{Setting the Stage}

Understanding electroweak symmetry breaking is arguably the biggest
challenge in particle physics, and we are confident that we will solve
it at the LHC. In the Standard Model, a single Higgs doublet gives
mass to the up--type and down--type quarks and also manifests itself
as one scalar Higgs boson. Electroweak precision data indicates that
this Higgs boson is light~\cite{lep_ew}. In the supersymmetric
extension of the Standard Model we need two Higgs doublets to give
mass to the up--type and down--type quarks and to cancel anomalies
of the fermionic partners of the Higgs bosons. The particle content of
a two-Higgs doublet model (2HDM) consists of two scalars, a
pseudo-scalar and a charged Higgs boson. It has been shown that the
LHC is guaranteed to find one supersymmetric Higgs boson in the weak
boson fusion production process with subsequent decay to tau
leptons~\cite{nolose,intense}. However, if we want to test a
supersymmetric Higgs sector the charged Higgs boson becomes
crucial. While there are many ways to add scalars to the Standard
Model spectrum and allow them to mix with the Higgs boson, for
example radions~\cite{radion}, a charged Higgs boson with all the
appropriate couplings as in the 2HDM is much harder to fake.

\subsection{Charged Higgs Bosons at the LHC}

Over the years, there have been many studies of charged Higgs bosons
at the LHC. The dominant production processes are in association with a top
quark~\cite{bg_th_pheno,me_1,me_2,bg_th_nlo}, in association with a
$W$ boson~\cite{gg_wh,bb_wh}, and charged Higgs pair
production~\cite{charged_pair_dy,charged_pair_gg, kniehl,obrian,
charged_pair_bb,charged_pair_wbf, charged_pair_bb_ex}. The most
promising decay channels are (depending on the charged Higgs mass)
$\tau\bar{\nu}$~\cite{pheno_dec_tau,atlas_dec_tau,cms_dec_tau}, 
$b\bar{t}$~\cite{pheno_dec_top,atlas_dec_top,cms_dec_top}, and
$Wh^0$~\cite{pheno_dec_wh}. According to studies by
ATLAS~\cite{atlas_dec_tau} and CMS~\cite{cms_dec_tau}, the process $pp
\to tH^- \to t(\tau \bar{\nu})$ is currently the most promising
combination, in particular for large values of $\tan\beta$. The reason
is that the Yukawa coupling of the charged Higgs boson to quarks
includes a term proportional to $m_b \tan\beta$, which means that the
$tH$ production cross section is enhanced by a factor $\tan^2
\beta$. Unfortunately, the same studies indicate that for $m_H > m_t$
the $tH$ production channel fails for $\tan\beta \lesssim 10$ because
of the reduced production rate.\smallskip

Pair production of charged Higgs bosons is particularly interesting
because three different production processes contribute to the same
final state: the usual Drell-Yan production process $q\bar{q} \to \hh$
through an $s$-channel $Z$ or a photon~\cite{charged_pair_dy}, the
loop--induced gluon fusion process $gg \to
\hh$~\cite{charged_pair_gg,kniehl,obrian}, and bottom--parton
scattering $b\bb\to \hh$. The latter two are enhanced as $\sigma
\propto \tan^4 \beta$ for large values of $\tan\beta$, which can
compensate for the loop factor suppression or the small bottom parton
density. The idea of this paper is twofold. First, in a detailed study
(including a next-to-leading order (NLO) calculation of the total cross 
section) of the
process $b\bb \to \hh$, we establish a reliable prediction for its
rate. Second, we show how the different production cross
sections can be added and how they contribute in the different areas
of 2HDM parameter space. In a brief addendum we will estimate the
supersymmetric QCD corrections to the process $b\bb \to \hh$ and test
the reliability of the approximation for the complete set of
supersymmetric diagrams by the leading (resummed) $\tan\beta$
contributions.\bigskip

\subsection{Bottom Parton Picture}
\label{sec:bottom_intro}

\begin{figure}[t]
 \includegraphics[width=15cm]{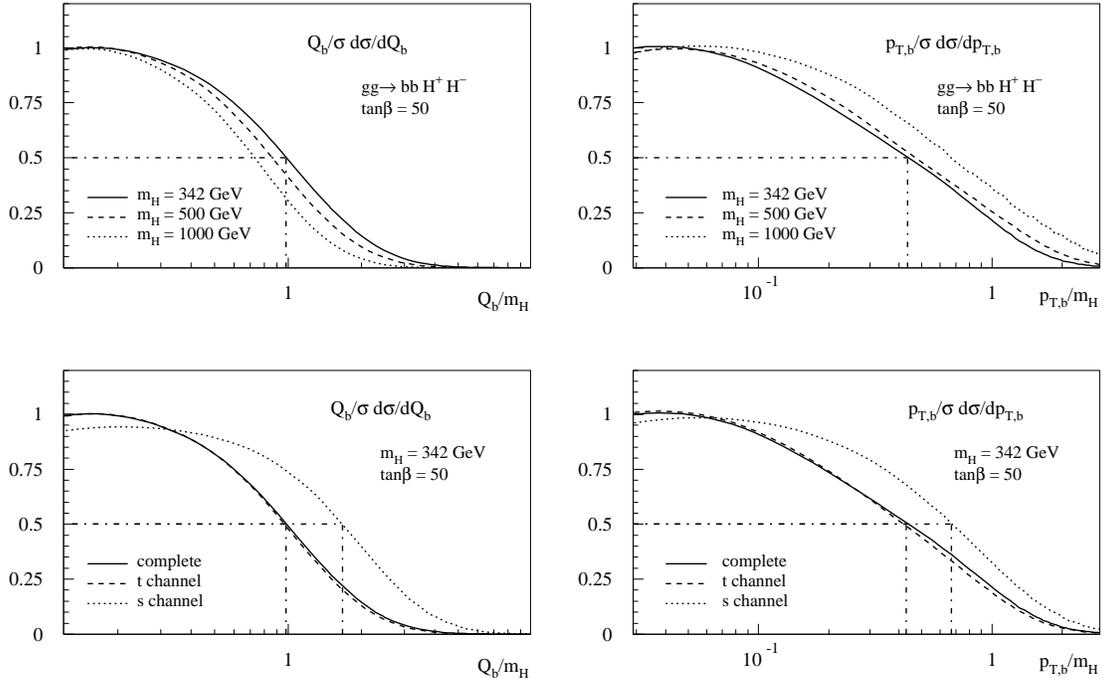}
 \caption[]{The dependence of the bottom--exclusive process $gg \to
 b\bb \hh$ on the bottom virtuality and transverse momentum. The upper
 set of figures show the variation with charged Higgs mass, while
 the lower figures show the distributions for the different contributing
 subprocesses, namely a $t$-channel top quark and an $s$-channel $Z$
 boson and photon. The $s$-channel Higgs contributions are heavily
 suppressed for this choice of parameters. We see that the shoulders
 of the $p_{T,b}$ distributions extend to values around $m_H/2$ or
 $M/4$ if the mass of the final state is identified with $M \equiv 2
 m_H$~\cite{me_1,me_2}.\label{fig:plateau}}
\end{figure}

The $\tan\beta$ enhancement of the charged Higgs Yukawa coupling,
which consists of one term proportional to $m_b \tan\beta$ and another
one proportional to $m_t \cot\beta$, immediately directs us to consider
processes with bottom quarks in the initial state. Conceptually, this
leads to the proper description of bottom partons. Lacking a
measurement of bottom quarks in the proton, we calculate a bottom
parton density from the gluon density, with gluons splitting into a
bottom quark pair. Because the bottom mass is much larger than
$\Lambda_{\rm QCD}$, this splitting can be described
perturbatively. It involves a large logarithm, which can be resummed
using DGLAP evolution~\cite{bottom_parton}.

We illustrate the bottom--parton approach using the process $bg \to
tH^-$~\cite{me_1,me_2,me_eduard}, but the arguments given in this
section are completely general. First, we compute the total cross
section for the gluon--induced process (\eg $gg \to \bb t H^-$),
integrating over the final state bottom quark. This class of processes
with explicit gluon splitting into bottom quarks, and with a bottom
jet in the final state, we refer to as bottom--exclusive. In the limit
of vanishing $m_b$, the differential cross section with respect to the
transverse momentum of the bottom quark will diverge like $d \sigma/d
p_{T,b} \sim 1/p_{T,b}$. The cross section after integrating over the
phase space of the outgoing bottom quark will (again for vanishing
$m_b$) be approximately proportional to $\log (p_{T,b}^{\rm
max}/p_{T,b}^{\rm min})$. In the presence of a finite bottom mass,
$m_b$ regularizes the cross section and yields a dependence $\log
(p_{T,b}^{\rm max}/m_b)$. If the upper integration boundary is given by
some hard scale in the process, for example the masses of some heavy
final state particles $M = m_H+m_t \gg m_b$, the total cross section
after integrating over the bottom jet will be proportional to a large
logarithm $\log M/m_b$. This logarithm can be resummed: switching from
the bottom--exclusive process $gg \to \bb t H^-$ to the bottom--parton
induced (bottom--inclusive) process $bg \to t H^-$ does precisely
that. From the above argument it is obvious that the bottom parton
density as a function of the bottom factorization scale grows roughly
like the leading logarithmic contribution $\log \mu_{F,b}$.

Historically, bottom--parton induced processes overestimated
the total cross section if compared to their exclusive counterpart. 
The reason was that the bottom--parton description introduces a
new parameter, namely the factorization scale of the bottom parton. It
corresponds to the maximum transverse momentum of the final state
bottom quark which is included in the bottom parton density. Since the
bottom density is currently perturbatively calculated we can estimate
the value of $\mu_{F,b}$, because it is implicitly part of the
definition of the bottom partons: it is the maximum value of $p_{T,b}$
in the bottom--exclusive process (again \eg $gg \to \bb t H^-$) for
which the differential cross section $d \sigma/d p_{T,b}$ behaves like
$1/p_{T,b}$ and can therefore contribute to the bottom parton density
--- or in terms of the discussion in the previous paragraph:
$\mu_{F,b} \equiv p_{T,b}^{\rm max}$. Following a naive dimensional
argument and identifying $\mu_{F,b} = M$ implicitly assumes that the
asymptotic behavior of the bottom--exclusive process extends to
$p_{T,b}^{\rm max}=M$. If for some reason $p_{T,b}^{\rm max} < M$ this
will automatically lead to an overestimate of the total
bottom--inclusive cross section.\smallskip

In previous work~\cite{me_1,me_2,me_eduard} we showed that indeed at the LHC it is more
appropriate to choose a smaller scale, $\mu_{F,b} \equiv p_{T,b}^{\rm max} \sim M/4$,
where $M$ is the hard scale in the process, \ie the mass of the heavy
states $M_X$ in the process $gg \to \bb X$ or $gg \to b\bb
X$. This scale choice can be derived in
general from the kinematics of the exclusive process at the
LHC~\cite{me_eduard}. Higher order calculations of the processes $b\bb
\to H$~\cite{bb_h_nnlo} and $bg \to bH$~\cite{bg_bh_nlo} have shown
that this choice of factorization scale is supported by the
stability of the perturbative series in $\alpha_s$. The perturbative
behavior of the total cross section $bg \to tH^-$ also confirms this
choice of scales. Moreover, a comparison between the NLO results for
the exclusive process $gg \to b\bb H$~\cite{gg_bbh_nlo} and the one--
or two--bottom inclusive processes $bg \to bH$ and $b\bb \to H$ 
showed that the different results are now in good agreement~\cite{bg_bh_nlo}. 
Note that the different cross sections
should not agree perfectly, as they include contributions at different orders in
perturbation theory. Because they differ mostly by the fact that the
bottom--parton induced processes resum the logarithm $\log
(p_{T,b}^{\rm max}/m_b)$, the agreement should be better for small final
state masses, in our case small $m_H$. On the other hand, they should
agree within their perturbative error bands, because we know the
difference between the calculations and we can estimate their
respective theoretical errors.

Going beyond total cross sections, we showed in Ref.~\cite{me_2} that for the process
$bg \to tH^-$, the distributions of the heavy final state particles are
not affected by the approximations of the bottom parton
picture, namely the assumption that the incoming bottom quarks are
massless and that they are collinear. These issues will be addressed
for $b\bb \to \hh$ in
section~\ref{sec:bottom_check}.\medskip

Before we analyze bottom--induced charged Higgs pair
production, we perform a consistency check. Based on general
kinematical arguments, it was shown that for LHC production
processes the bottom factorization scale should be chosen around $M/4
= m_H/2$. We repeat the numerical analyses of
Refs.~\cite{me_1,bg_bh_nlo,me_2} and show the results on which this
argument is based in Fig.~\ref{fig:plateau}. The two left panels show
the intermediate bottom virtuality, the right pair show the transverse
momentum of the intermediate or final state bottom. In accordance with
earlier results we see that the shoulder limiting the $1/Q_b$ or
$1/p_{T,b}$ behavior for large momentum transfer sits at smaller
values for the transverse momentum than for the virtuality. For
different charged Higgs masses both shoulders scale with the mass,
which is equivalent to the claim that the bottom factorization scale
has to be proportional to the hard scale $\mu_{F,b} \propto M = 2
m_H$. The smaller maximum values of $Q_b$ and
$p_{T,b}$ for larger values of the charged Higgs mass is indicative of
the limited phase space at the LHC. The value for $p_{T,b}^{\rm max}$,
up to which we can observe something like a $1/p_{T,b}$ behavior, is
clearly smaller than $M= 2 m_H$ and in agreement with the general
estimate $\mu_{F,b} \sim M/4 = m_H/2$. Also in Fig.~\ref{fig:plateau}
we see that the shoulder for the $s$-channel process extends to larger
values of $Q_b$ and $p_{T,b}$ by about a factor of 1.5. This can be
understood from the approximate determination of the factorization
scale in Ref.~\cite{me_eduard}: the position of the shoulders in the
$Q_b$ and the $p_{T,b}$ curves can be approximately computed relying 
only on the parton $x$ behavior of the gluon density. However, the
results $Q_b^{\rm max} \sim M/2$ and $p_{T,b}^{\rm max} \sim M/4$
assume that the heavy final state is produced at threshold. The
difference between the $s$-channel and $t$-channel processes is that
$t$-channel Higgs pair production can proceed through an $S$ wave,
which cross section increases proportional to the Higgs velocity
$\beta$~\footnote{Note that this $\beta$ has nothing to do with
$\tan\beta$, it is defined as $\beta = \sqrt{1-4
m_H^2/\hat{s}}$. Since both definitions are very common and there is
little chance to confuse them we prefer to stick to this notation.}
at threshold, while the $s$-channel process increases like $\beta^3$
and therefore shows a delayed onset. The $s$-channel process is
strongly suppressed at threshold and produces the bulk of the Higgs
pairs at a larger partonic center--of--mass energy (we show the corresponding
$p_{T,H}$ distributions in sections~\ref{sec:sigma} and
\ref{sec:combo}). Thus, for the $s$-channel process, the
typical final state invariant mass is larger, which means 
effectively $M>2 m_H$. This is what we observe for the gauge boson
exchange process in Fig.~\ref{fig:plateau}. We also notice that the
determination of the bottom factorization scale has a theoretical
error, because we have to approximate the physical curves with a
box-shaped object. Effects like threshold suppression of the
$s$-channel diagrams are covered by the theoretical error, and we
discuss the different scale variations in detail in
section~\ref{sec:sigma}.

\subsection*{Organization of the Paper and Conventions}

The paper is organized as follows: in this section we
reviewed the status of charged Higgs searches at the LHC and
the status of the bottom--parton approach. In section~\ref{sec:sigma}
we compute the bottom--parton induced cross section to NLO. We
show how the description is perturbatively stable with respect to
QCD corrections. In section~\ref{sec:bottom_check} we use the NLO
predictions for the distributions of the heavy final states to prove
the validity of the bottom--parton approach. The results from these two
sections allow us to quantitatively compare the different
production processes for charged Higgs pairs at the LHC in
section~\ref{sec:combo}. This comparison covers total rates as well as
differential cross sections. As an addendum we compute the
full SUSY-QCD corrections to the bottom--parton scattering process in
section~\ref{sec:susy}.\bigskip

Before we proceed with the discussion of the results we briefly give
all conventions used for our calculations.
All bottom--parton induced cross sections are computed with
zero bottom mass. When we check our results using final state bottom
quarks we use the on-shell mass $m_b = 4.6$~GeV, or for testing
purposes a second value 0.46~GeV. The top quark mass is always chosen
as $m_t = 178$~GeV. In contrast, the bottom and top Yukawa couplings
are defined in the \msbar scheme, \ie numerically we use $m_b(\mu_R)$
and $m_t(\mu_R)$. Unless explicitly mentioned we use the renormalization
scale $\mu_R = m_H$ and the bottom factorization scale $\mu_{F,b} =
m_H/2$. The parton densities are CTEQ6~\cite{cteq6}, consistently chosen
to be the LO or NLO sets L1 or M; $\alpha_s$ is also calculated at
equivalent order.

\section{Bottom--Induced Production Process}
\label{sec:sigma}

\begin{figure}[t]
 \includegraphics[width=15cm]{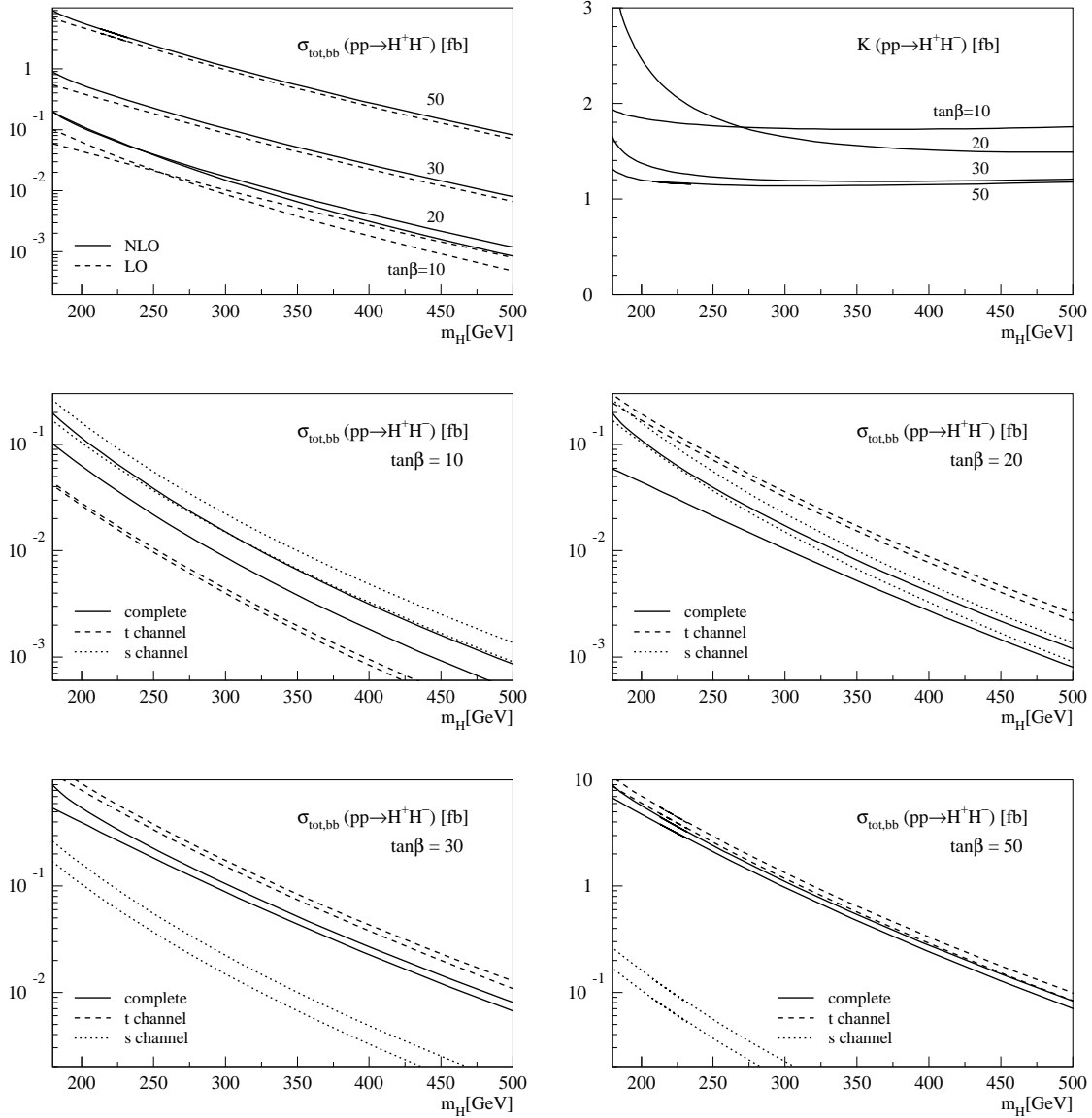}
 \caption[]{LO and NLO total cross section for the process
 $b\bb \to \hh$ at the LHC. The $K$ factor is defined consistently as
 $\sigma_{\rm NLO}/\sigma_{\rm LO}$. The parameters in the 2HDM Higgs
 sector are chosen consistently for each value of $\tan\beta$ and
 charged Higgs mass $m_H$. In the lower two rows the contributions
 from $s$-channel Higgs and $\gamma,Z$ exchange and $t$-channel top 
 quark exchange are shown. In those figures the NLO curve is
 always the upper one in each pair.\label{fig:sig_nlo}}
\end{figure}

\begin{figure}[t]
 \includegraphics[width=15cm]{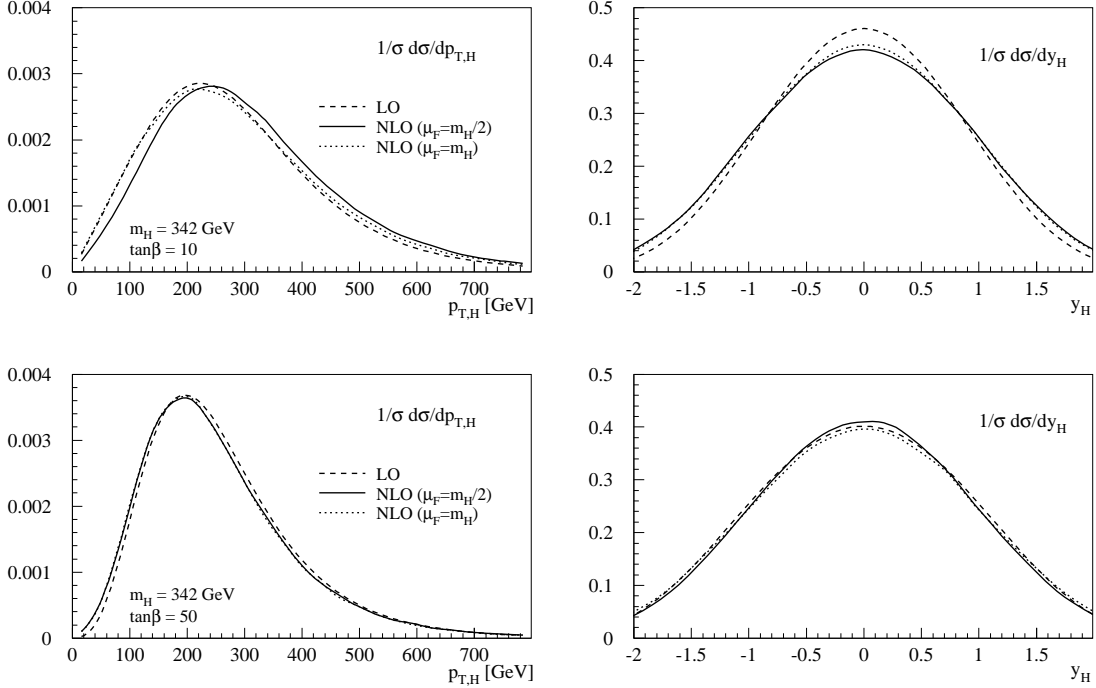}
 \caption[]{The rapidity and transverse momentum distributions at
 LO and NLO. The results are shown for two different
 values of $\tan\beta$, so illustrate the behavior of the $s$-channel
 and the $t$-channel diagrams. We include a set of NLO curves for the
 factorization scale $\mu_F = 2\mu_F^0 = m_H$ to illustrate the
 remaining theoretical uncertainty.\label{fig:distri}}
\end{figure}

Before we can discuss the different charged Higgs pair production 
processes at the LHC (in section~\ref{sec:combo}), we have to find a
way to reliably predict the bottom--induced production rate $b \bb \to
\hh$. This shortcoming was noticed in an earlier
paper~\cite{kniehl}, and with the recent progress in understanding
bottom--parton induced processes we can now fill this gap. The key
ingredient for the correct evaluation of production processes of the
kind $b \bb \to \hh$ is the choice of bottom parton factorization scale,
which we reviewed in detail in section~\ref{sec:bottom_intro}. Next,
we will compute the NLO corrections (for now in the 2HDM, \ie without
gluino contributions, which are added in section~\ref{sec:susy}).

We calculate the ${\cal O}(\alpha_s)$ corrections to the process
$pp/b\bb \to \hh +X$ following two different approaches. First, we use
a one-cutoff technique to calculate the total cross sections. These
results will be made publicly available as part of an extension to the
Prospino2 package~\cite{prospino2}. This approach has the advantage of
being numerically fast and reliable. In a second step we repeat the
calculation with two cutoffs for the soft ($\delta_s=10^{-4}$) and
collinear ($\delta_c = \delta_s/100$) divergences~\cite{brian}. This
allows us to calculate distributions for the heavy final state
particles. These distributions are a crucial check of the validity of
the bottom--parton approximation, which we use to improve the
prediction of the total cross section. Moreover, we will be able to
compare the differential cross sections for the different charged
Higgs pair production mechanisms in section~\ref{sec:combo}.\bigskip

\underline{NLO corrections:} We present the results of our NLO
calculation in Fig.~\ref{fig:sig_nlo}. The total cross sections are
plotted as a function of the charged Higgs mass for four different
values of $\tan\beta$. The 2HDM parameters are chosen
consistently~\cite{higgs_para}. Before discussing the higher-order
corrections we note that indeed the leading order (LO) cross section
increases like $\tan^4\beta$ for large values of $\tan\beta$. The $K$
factor in the second panel is defined as $\sigma_{\rm NLO}/\sigma_{\rm
LO}$. The size of the NLO correction is strongly dependent on the
Higgs mass and on the value of $\tan\beta$. To understand this we
separate the three different contributions to the production rate.

The bottom--induced Drell--Yan contribution proceeds through an
$s$-channel photon or $Z$ boson. The coupling to the photon is
independent of $\tan\beta$ and the dependence of the $Z$ coupling is
small for large enough values of the pseudoscalar and charged Higgs
masses (the decoupling limit). The NLO corrections for the
bottom--induced Drell--Yan process vary around $K=1.60 \cdots 1.55$
for Higgs masses from 160~GeV to 500~GeV. This correction is somewhat
larger than for the light--flavor Drell-Yan process, which is due to
the $x$ dependence of the bottom parton densities and can also be
observed in the perturbative behavior of the Standard Model $b\bb \to
H$ process~\cite{bb_h_nnlo}.

A second set of $s$-channel production processes is made possible by
the incoming bottom quarks and their finite Yukawa coupling: we can
couple the charged Higgs pair to $s$-channel neutral CP-even Higgs
bosons $h^0,H^0$. CP-odd pseudoscalar exchange is forbidden by the
CP symmetry of the final state. The Higgs self-couplings $h^0 \hh$ and
$H^0 \hh$ are strongly dependent on the Higgs sector~\cite{charged_pair_gg}
parameters, but the relative size of the Higgs
contribution compared to the Drell--Yan process is basically
negligible; the suppression varies from $10^{-2}$ for small Higgs
masses below 200~GeV to $10^{-3}$ for Higgs masses around
500~GeV. While it is
possible that there are regions of 2HDM parameter space where this
hierarchy is less pronounced, the impact of the $s$-channel Higgs
exchange is small for our analysis. We note, however, that the NLO
corrections to the $s$-channel Higgs exchange are smaller than for the
Drell--Yan process, namely $K=1.15 \cdots 1.20$ (for charged Higgs
masses from 160~GeV to 500~GeV). This is in part an effect of the
\msbar renormalization of the bottom Yukawa coupling. The combination
of the two $s$-channel processes receives a fairly constant NLO
correction, from $K=1.58$ for small Higgs masses and $\tan\beta=10$ to
$K=1.55$ for large values of $\tan\beta$ or large Higgs
masses.\smallskip

For the total rate the $s$-channel production processes compete with
$t$-channel top quark exchange. The $btH^+$ coupling consists of two
contributions, $y_b \tan\beta$ and $y_t \cot\beta$. Over the entire
parameter space $\tan\beta>10$, the $y_b^4$ contribution dominates
over the
$y_t^4$ contribution by at least one order of magnitude. We note that
because of the different running of the bottom and top Yukawa
couplings the $K$ factor of the $y_b^4$ contribution is considerably
smaller than for the $y_t^4$ contribution~\cite{me_1}. This leads to a
$\tan\beta$ dependence of the NLO corrections to the $t$-channel
process. For small Higgs masses the size of the correction varies as
$K=1.06 \cdots 1.36$ (for $\tan\beta=10$ to $50$). For larger values
of the Higgs masses (and thereby the renormalization scale) the behavior
of $K$ becomes flat.\smallskip

Combining the $s$-channel and $t$-channel contributions, we see
that they interfere destructively. For small Higgs masses the LO rates
are almost identical in size for $\tan\beta \sim 20$. Because the two
classes of diagrams have very different NLO corrections this leads to
huge relative QCD corrections to the total rate. We observe this in
Fig.~\ref{fig:sig_nlo}, where the $K$ factors are fairly constant for
the $s$-channel dominated regime $\tan\beta \lesssim 10$, and for the
$t$-channel dominated regime $\tan\beta \gtrsim 30$. The curve for
$\tan\beta=30$ shows a small remaining increase only for very small
Higgs masses. The slight difference between the large $m_H$ behavior
of the $\tan\beta=30$ and $\tan\beta=50$ curves is also due to the
interference between the $s$- and $t$-channel subprocesses.

We emphasize that in this paper we show results only for large charged
Higgs masses $m_H > m_t$. For small Higgs masses there is a technical
complication, because the production process $bg \to tH^-$ and its
charge conjugate will be part of the NLO corrections. We subtract the
corresponding divergence consistently in the small width
approximation, which is also used in
Refs.~\cite{me_2,prospino2,on_shell}. The public computer code in the
extension to Prospino2 will, however, include this region of parameter
space. Phenomenologically it is less interesting, because charged Higgs
bosons will be produced in large numbers in anomalous top quark decays.
Matching with the off-shell process $bg \to tH^-$ is studied in
detail in Ref.~\cite{me_2}.\bigskip

\underline{Distributions:} The effect of NLO corrections on the
distributions of the heavy final state Higgs bosons is shown in
Fig.~\ref{fig:distri}. The case $\tan\beta=50$ corresponds to the
parameter point SPS4~\cite{sps}, and the parameter choice $\tan\beta=10$
is the same in all parameters except for $\tan\beta$. According to
Fig.~\ref{fig:sig_nlo}, $t$-channel top quark exchange is effectively switched
off this way. All that is left is $s$-channel gauge boson exchange.  For
$\tan\beta=50$, the rate is completely dominated by $t$-channel top
exchange: the relative size of the two contributions, $\sigma_{s,\,
{\rm only}}/\sigma_{t,\, {\rm only}}$, varies from $2.0\%$ to $1.0\%$
at LO and from $2.5\%$ to $1.5\%$ at NLO (with a charged
Higgs mass between 200~GeV and 500~GeV).  While the shift in this
fraction is sizable, as we would expect from the different $K$
factors described above, any distribution for $\tan\beta=50$ will be
completely independent of the $s$-channel contributions. Similar to
higher order corrections to supersymmetric particle
production~\cite{prospino2}, we indeed see that a constant $K$ factor
in the ($p_{T,H}-y_H$) plane is a reasonable approximation within any
error estimate.\smallskip

The situation is slightly different for $\tan\beta=10$. Here the peak
in the $p_{T,H}$ distribution shifts by 20~GeV when we switch to the
slightly harder NLO description. The interference between the $s$
channel and the $t$ channel subprocesses plays no role ---
$\sigma_{t,\, {\rm only}}/\sigma_{s,\, {\rm only}}$ decreases from
typically 0.25 at LO to NLO values around 0.18, which we
checked is not sufficient to impact the distributions
visibly. Instead, the shift towards a slightly harder spectrum is a
result of the three--particle phase space at NLO. The extra jet
balances the slightly harder charged Higgs bosons in the final state. The
same effect actually appears for $\tan\beta=50$, but it is less
pronounced. To estimate how the shift in the $p_{T,H}$ distribution
compares with the theoretical uncertainty, and to understand the
difference in the size of the shift between $s$--channel and
$t$--channel diagrams, we change the bottom factorization scale to
twice its usual value: $\mu_F = m_H$. From
section~\ref{sec:bottom_intro} we know that this will make the
$s$--channel ($P$--wave) subprocess more similar to $t$--channel
($S$--wave) top quark exchange. Indeed, in Fig.~\ref{fig:distri} we 
see that the LO curve shows no visible effect, and that the NLO and
the LO distributions are now almost identical.

As we pointed out in section~\ref{sec:bottom_intro} the determination
of the bottom factorization scale shows an inherent theoretical error,
because it is not possible to tell where exactly the asymptotic
behavior of the bottom--exclusive processes is no longer valid. The
observed scale-dependent shift between the LO and NLO
distributions reflects this remaining theoretical uncertainty, and
shows that it is perturbatively under control.\bigskip

\begin{figure}[t]
 \includegraphics[width=14cm]{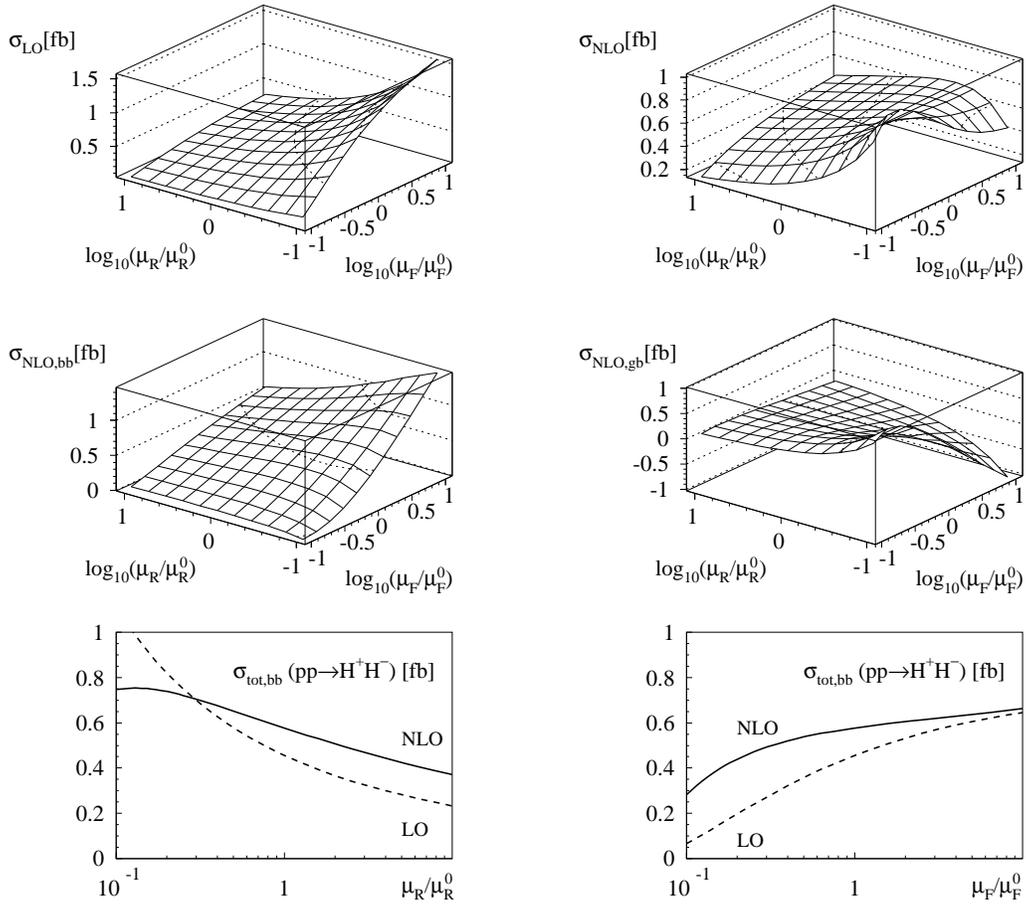}
 \caption[]{Factorization and renormalization scale dependence of the
 total cross section. The scales are varied around the central values
 $\mu_F^0 = m_H/2$ and $\mu_R^0 = m_H$. In the second row the
 contributions to the NLO rate are separated by their different
 initial state. In the third row the scale, which is not varied, is
 fixed to its respective value $\mu^0$. The LO curves are the dashed ones
 and the NLO curves are the solid lines. For the charged Higgs mass we
 use our central parameter point SPS4 with $\tan\beta=50$ and
 $m_H=342$~GeV.\label{fig:scale}}
\end{figure}

\underline{Scale dependence:} Even though we have argued before that
the bottom factorization scale should be chosen around $m_H/2$, the
renormalization and factorization scales are nevertheless free
parameters introduced by the perturbative expansion of the hadronic
production cross sections. The bottom factorization scale plays a
special role, because it is not actually extracted from data. Instead,
it is calculated from the measured gluon density in a perturbative
regime. However, while we can claim what the (roughly) appropriate
values for $\mu_{F,b}$ should be, it is also obvious from
Fig.~\ref{fig:plateau} that a serious theoretical error margin should
be assigned to the determination of the appropriate scale. We
therefore proceed to check the factorization and renormalization scale
dependence of the cross section $pp \to \hh$ at LO and 
NLO. Because the scales are artifacts of perturbation theory, we
expect the scale dependence to be a measure of the theoretical error
on the total cross section, and we expect it to flatten after we
include the NLO contributions.\smallskip

The scale dependences of the LO and NLO cross sections are
shown in Fig.~\ref{fig:scale}. For the two dimensional plots both
scales are varied up to factors 1/10 and 10, respectively, around
their central values $\mu_F^0=m_H/2$ and $\mu_R^0=m_H$. The central
value of the factorization scale was motivated in
section~\ref{sec:bottom_intro}, and the central value of the
renormalization scale is motivated by the production of heavy
supersymmetric particles at the LHC~\cite{prospino2} and by the
production of Higgs bosons from bottom--parton initial
states~\cite{me_1,me_2,bb_h_nnlo}. Note that the renormalization scale
covers the running top and bottom Yukawa couplings at LO,
whereas the strong coupling $\alpha_s$ only enters at NLO.

From the two-dimensional behavior shown in Fig.~\ref{fig:scale}, we see
that the NLO rate has a considerably flatter scale dependence than its
LO counterpart. There is, however, one exception to this
rule: for very small $\mu_F$ and very small $\mu_R$ the perturbative
expansion of the process $b \bb \to \hh$ becomes unstable. This is the
regime where the bottom--parton description breaks down: a small bottom
factorization scale means that hardly any phase space of the gluon
splitting in the bottom--exclusive process is resummed and very little
of the bottom--exclusive production rate is included in the bottom
parton density. This means that the $bg$--induced process which
contributes to the NLO rate will dominate over the $b\bb$--induced
LO process. This effect becomes more pronounced if the
perturbative expansion parameter which suppressed the $bg$--induced
channel as compared the $b\bb$--induced Born process is large, \ie for
a small renormalization scale which increases the value of
$\alpha_s(\mu_R)$. The same feature can be seen in the associated
production of a charged Higgs with a top quark~\cite{me_2}.\smallskip

If we stay away from this extreme (and unphysical) scale choice, we
can attempt to estimate the theoretical error of the LO and
NLO total cross sections for $pp/b\bb \to \hh+X$. A reasonable
range of scales could be a band from $\mu^0/3$ to $3 \mu^0$ around the
central values. First, we see that the LO and NLO error
bands for both scales would overlap in Fig~\ref{fig:scale}. This is
not crucial, but it is certainly a welcome feature, which indicates
that there are no large constant terms entering the cross section
calculation at the NLO level. Moreover, we see that the factorization
scale dependence covers a little less than a $20\%$ error band, and
the renormalization scale points to an error slightly above
$10\%$. Even though the two scale dependences cancel each other if we
identify $\mu_R/\mu_R^0 \sim \mu_F/\mu_F^0$ we prefer to
conservatively estimate the overall error by adding the two errors in
square. This leads us to an estimate of the theoretical error (due to
higher order corrections) for the NLO total cross section prediction
of around $25\%$. We emphasize that this estimate cannot be a precise
error band with a statistical meaning, as for example statistical or
systematic experimental errors are. The range of scales we choose to
cover is not fixed by first principles, and it is by no means
guaranteed that no large finite terms come in at the NNLO
level. However, we know that we have resummed the leading large
logarithms into the bottom parton densities, and we see a well behaved
perturbative behavior going from the LO to the NLO cross
section calculation.

\section{Bottom Parton Approximation}
\label{sec:bottom_check}

\begin{figure}[t]
 \includegraphics[width=15cm]{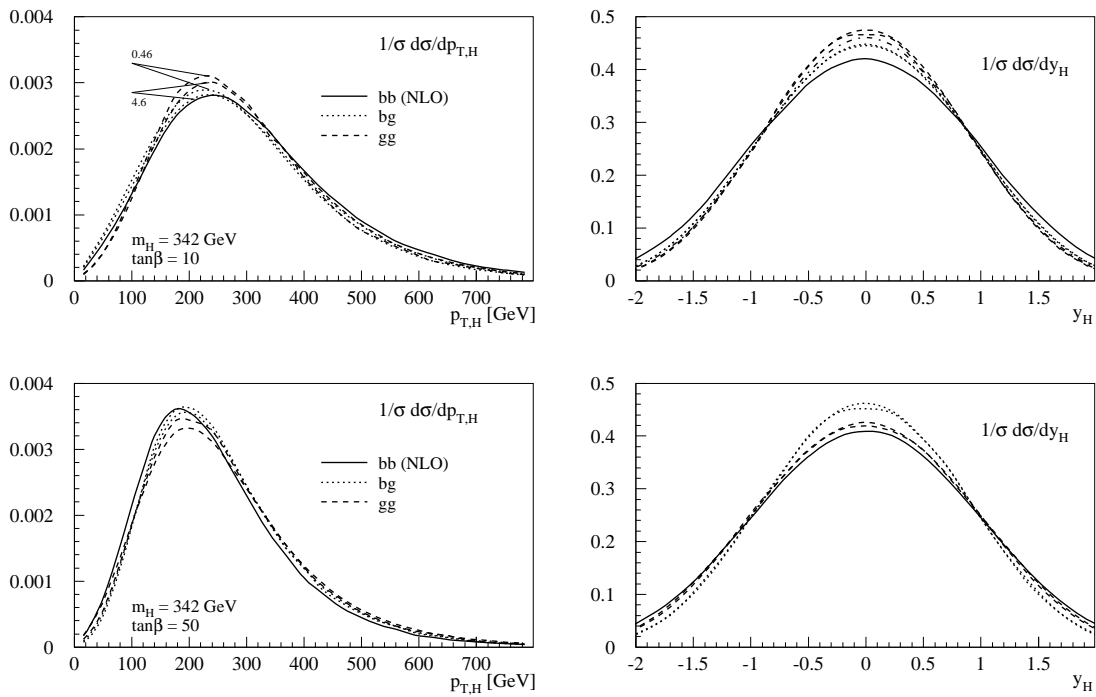}
 \caption[]{The rapidity and transverse momentum distributions of the
 final state charged Higgs boson for different processes. The two-bottom and
 one--bottom exclusive results (with the initial states $gg$ and $bg$)
 are given for the physical mass $m_b=4.6$~GeV and for a hypothetical value
 of $0.46$~GeV. The NLO results for the $b\bb$-initiated process is shown
 for massless bottom quarks.\label{fig:test_mb}}
\end{figure}

The bottom--parton approach which we use to compute the double--bottom
inclusive process $b\bb \to \hh$ resums large logarithms and improves
the prediction of the bottom--inclusive cross sections, but it also
implicitly makes two approximations. One is that incoming bottoms are
massless, the other one is that the bottom jets over which we
integrate are collinear. For total cross sections these approximations
are well founded: bottom mass effects are suppressed by powers of
$m_b/m_H$, so they are well below any errors we quote on the NLO total
cross section calculation. The kinematical configuration of the final
state does not enter the computation of the total cross section, which
means that we do not expect any non-negligible finite-$p_{T,b}$
effects for the total cross section calculation using bottom
partons.\medskip

In section~\ref{sec:bottom_intro} we showed that the bottom--parton
approach works well for total cross sections. In some sense this is by
construction, because we estimate the value of the only new parameter, the bottom
factorization scale, such that the total cross section prediction is
consistent with the definition of the bottom parton densities as they
are computed from the bottom--exclusive process. The only serious test
of the bottom--parton approach is the check that our estimate of the
factorization scale indeed predicts $\mu_{F,b} \propto M$, where $M$
is the hard scale in the process, usually chosen as the sum of
final state masses. Differential cross sections, in contrast,
are a non-trivial check of our approach. Phenomenologically
it is crucial that the distributions of the heavy final states are not
significantly affected by the two approximations mentioned above
(here, significantly means with a larger effect than the perturbative
error of the NLO calculation, which we estimated to be around $25\%$ for the
total cross section).\smallskip

\underline{Massless bottom quark approximation:} The first approximation we
have to check is the assumption that the bottom partons and bottom
jets involved are assumed to be massless. This is easy to estimate,
because we can compute the one--bottom or two--bottom exclusive
processes at tree level with different bottom masses. There is a
technical complication in the one--bottom exclusive processes $bg \to
b\hh$, where we have to choose the incoming bottom as massless and the
outgoing bottom as massive. Because of SU(3) gauge invariance we
cannot couple a massless bottom and a massive bottom to a
gluon. However, in our $s$ and $t$-channel diagrams we are not
sensitive to U(1) or SU(2) gauge invariance, so we can switch from
massless incoming bottoms to massive outgoing bottoms at the $bbZ$ and
$bb\gamma$ vertices. 

We compute all distributions for the physical on-shell mass
$m_b=4.6$~GeV and for a hypothetical $0.46$~GeV. Apart from the NLO
predictions for the bottom--inclusive process $b\bb \to \hh$, the
normalized distributions for the rapidity and the transverse momentum
of the final state Higgs bosons are computed for the processes $gg \to
b \bb \hh$ and $bg \to b \hh$. These five curves are shown in
Fig.~\ref{fig:test_mb}. Because the $s$-channel neutral Higgs bosons do
not contribute visibly, one can think of the different values of
$\tan\beta$ as a rescaling of the independent Yukawa couplings; the
parameter choice $\tan\beta=10$ switches off $t$-channel top quark
exchange and leaves us with $s$-channel gauge boson exchange.

In Fig.~\ref{fig:test_mb} we see that the bands defined by varying the
bottom mass for a given process are considerably smaller than the
difference between the different initial states, \ie the difference
between the bottom--induced and gluon--induced processes. We also see
that for smaller values of $m_b$ the Higgs distributions become
slightly softer. While naively the finite phase space would move the
$p_{T,H}$ distributions to a softer regime when we increase the bottom
mass, the opposite happens: a
smaller bottom mass in the final state jet cuts off the $p_{T,b}$
distributions at smaller values of $p_{T,b}$. Because for both 
bottom--exclusive processes the charged Higgs bosons balance the final state
bottom jet in the transverse plane, the $p_T$ spectrum for all
particles becomes softer when the bottom jets become softer, which is
the case for a smaller cutoff value $p_{T,b}^{\rm min} \sim
m_b$. However, this effect is already part of the collinear bottom
quark approximation, which we consider next.
Therefore, at this point we conclude that bottom quark mass
effects are negligible compared to the collinear bottom quark 
approximation.\medskip

\begin{figure}[t]
 \includegraphics[width=15cm]{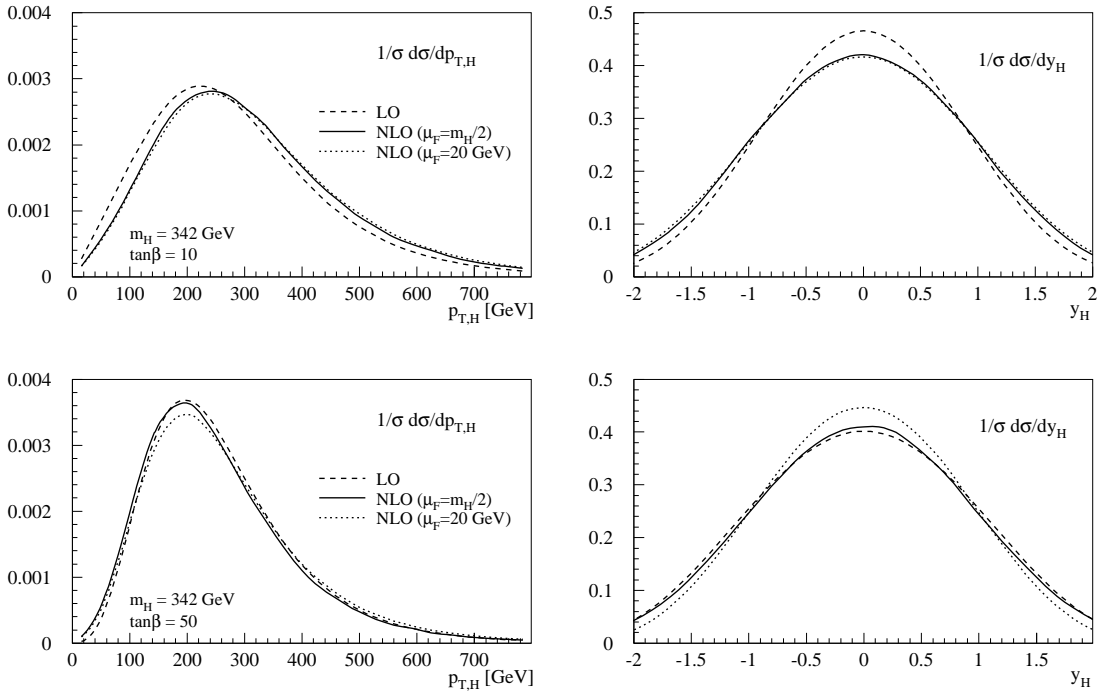}
 \caption[]{The rapidity and transverse momentum distributions for the
 final state charged Higgs boson for different choices of the factorization
 scale. The small scale choice $\mu_{F,b} = 20\GeV$ interpolates into
 the region where the formally-NLO subprocess $bg \to b \hh$ dominates
 the NLO rate.\label{fig:test_coll}}
\end{figure}

\underline{Collinear approximation:} From the above considerations we
know that the dominant error is the assumption that the final state
bottom jets are collinear and are resummed into the bottom
parton densities. We can test the validity of this assumption using
the NLO differential cross sections for the process $b\bb \to \hh$:
part of the NLO corrections are real jet emission diagrams, either
real gluon emission from the Born process or the bottom quark
emission $bg \to b\hh$, which we usually refer to as crossed
diagrams. In Fig.~\ref{fig:scale} we saw (and discussed in
detail) that, in the limit of $\mu_F \to m_b$, the perturbative
expansion around the LO process $b\bb \to \hh$ fails. The
bottom parton density becomes small, because it is defined to include
the splitting of the gluon into a bottom pair up to $p_{T,b} <
\mu_F$. With the lower cutoff $p_{T,b}^{\rm min} \sim m_b$ this means
that less gluon splitting is included in the bottom
parton density and more of it has to be considered explicitly, which 
in turn means accounting for the gluon-initiated process
$bg \to b \hh$, and eventually $gg \to b\bb\hh$. As we have seen, the
$K$ factor for the process $b\bb \to \hh$ blows up in this limit,
because already the NLO contribution $bg \to b \hh$ becomes dominant.

This behavior illustrates how the variation of $\mu_F$ interpolates
consistently between two regimes: for large $\mu_F$ the LO
process $b\bb \to \hh$ is dominant. In fact, for
$\mu_{F,b} = m_H/2$ it is designed precisely to minimize the
effect of the NLO contribution $bg \to b
\hh$~\cite{me_1,me_eduard}. For smaller $\mu_F$ the formerly NLO
contribution $bg \to b\hh$ will become dominant. The NLO calculation
of $b\bb \to \hh$ combines these two processes consistently (using 
the zero bottom quark mass approximation, which we tested
above)~\cite{me_1,me_2}. We should note that a NNLO calculation
including the $gg \to b\bb \hh$ subprocess would of course be
perfectly suited for this test, but goes beyond the scope of our
paper. Moreover, we emphasize that if the bottom--parton approach using
the collinear bottom quark approximation should not work well, this behavior
will already show up in the NLO interpolation.\medskip

In Fig.~\ref{fig:test_coll} we show the numerical results of this
comparison. As usual, we test the assumption for two values of
$\tan\beta$ or in other words independently for the $s$-channel and
$t$-channel subprocesses. All shifts from LO to NLO curves
were discussed in section~\ref{sec:sigma}.  To test the
bottom--parton approximation we now change the factorization scale to
$\mu_F = 20$~GeV. We see that for both the $s$-channel and
$t$-channel processes the $p_T$ distributions become slightly harder
when we compare the leading order and the low--scale curves. The 
now-dominant three particle kinematics is responsible for this shift.  
In an earlier work we showed that part of this effect is also due to
the hardening gluon spectrum when we move to smaller values of $\mu_F
\equiv \mu_{F,g}$ in the bottom--gluon initial state~\cite{me_2}. We
note, however, that the shifts in the $p_{T,H}$ and $y_H$
distributions are well within the perturbative uncertainty. As we
discussed in section~\ref{sec:sigma}, the NLO distributions for the
central factorization scale $\mu_F=m_H/2$ have to interpolate between
the LO and the low--scale curves, because they are affected by the two-
and three-particle kinematics. It is a curious coincidence
that the NLO rapidity distributions for this scale choice are much
more similar between the $s$- and $t$ channels than any of the other
cases. We therefore conclude that the collinear approximation in the
bottom--parton picture is indeed valid for the distributions of the
(heavy) final states. We have made similar checks for larger Higgs
masses, and both the zero-mass and collinear approximations
improve in accuracy. This is in agreement of what one would expect
from the bottom--parton approach~\cite{bg_bh_nlo}.


\section{Combination of Production Processes}
\label{sec:combo}

\begin{figure}[t]
 \includegraphics[width=15cm]{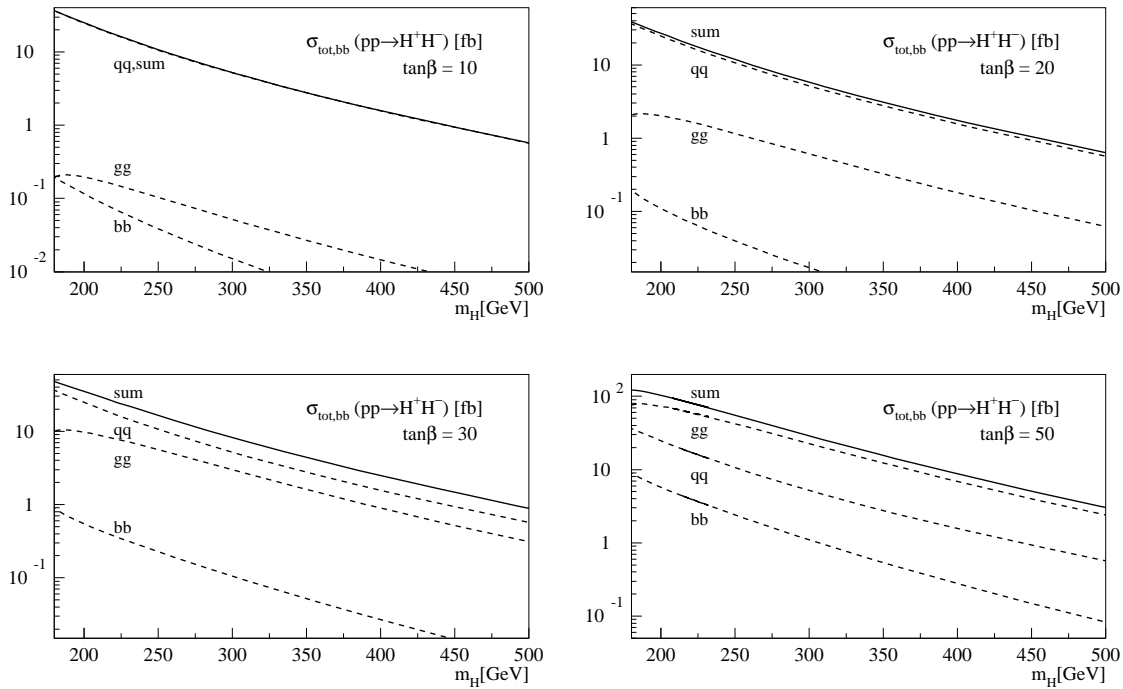}
 \caption[]{Production cross sections for pair production of
 charged Higgs bosons. The curves for the bottom--induced process (sum
 of $s$- and $t$-channel) is labeled $bb$. The Drell--Yan process with
 only light-flavor incoming quarks is labeled $DY$, and the gluon
 fusion process through the one--loop amplitude is labeled $gg$. The
 cross sections are given for different values of $\tan\beta$. The
 bottom--induced production rate and the Drell--Yan production rates
 are computed at NLO.\label{fig:sig_all}}
\end{figure}

Charged Higgs boson pairs can be produced in three different ways at
the LHC~\cite{kniehl}. Theoretically the most straight-forward
production process is light--flavor quark scattering via an
$s$-channel photon and $Z$~\cite{charged_pair_dy}. As described in
section~\ref{sec:sigma}, the charged Higgs couplings are nearly
independent of $\tan\beta$, as long as the pseudoscalar Higgs mass and
the charged Higgs mass are large enough. We calculate the NLO
production cross section in complete analogy to the production of two
supersymmetric sleptons, a production process which is part of the
publicly available computer program Prospino2~\cite{prospino2}. The
factorization and renormalization scales are chosen as $m_H$. In
Fig.~\ref{fig:sig_all} we see that for charged Higgs masses between
160 and 500 GeV the NLO rate Drell--Yan process varies between 58~fb
and 0.23~fb. This rate is computed without any cuts on the final state
particles. The size of the NLO corrections for this process varies
from $K=1.27$ to $K=1.17$ (for Higgs masses from 160~GeV to
500~GeV). The remaining theoretical error on the NLO cross section can
be estimated to be less than $25\%$~\cite{prospino2}. For small values of
$\tan\beta \lesssim 20$ this production process is the dominant
channel at the LHC.\smallskip

Similar to the light--flavor Drell--Yan process, charged Higgs pairs
can also be produced in bottom--parton scattering. We analyzed this
production process in detail in this paper and can now quote a
reliable estimate of the production cross section. We find that the
$s$-channel and the $t$-channel subprocesses interfere
destructively. The LO total cross section is enhanced by
$\tan^4\beta$ for large values of $\tan\beta$. In
Fig.~\ref{fig:sig_all} we see that even for $\tan\beta = 50$ the NLO
cross section is still below 10~fb. In this region of parameter space
the $t$-channel subprocess dominates completely. Compared to earlier
analyses we find that we have to use running \msbar Yukawa couplings
and a low bottom factorization scale, which suppresses the cross
section to a level where it will always stay below the light--flavor
induced rate as well as below the gluon--induced process discussed in
the next paragraph. The remaining theoretical error on this total
cross section we estimated in section~\ref{sec:sigma} to be on the
order of $25\%$.\smallskip

\begin{figure}[t]
 \includegraphics[width=15cm]{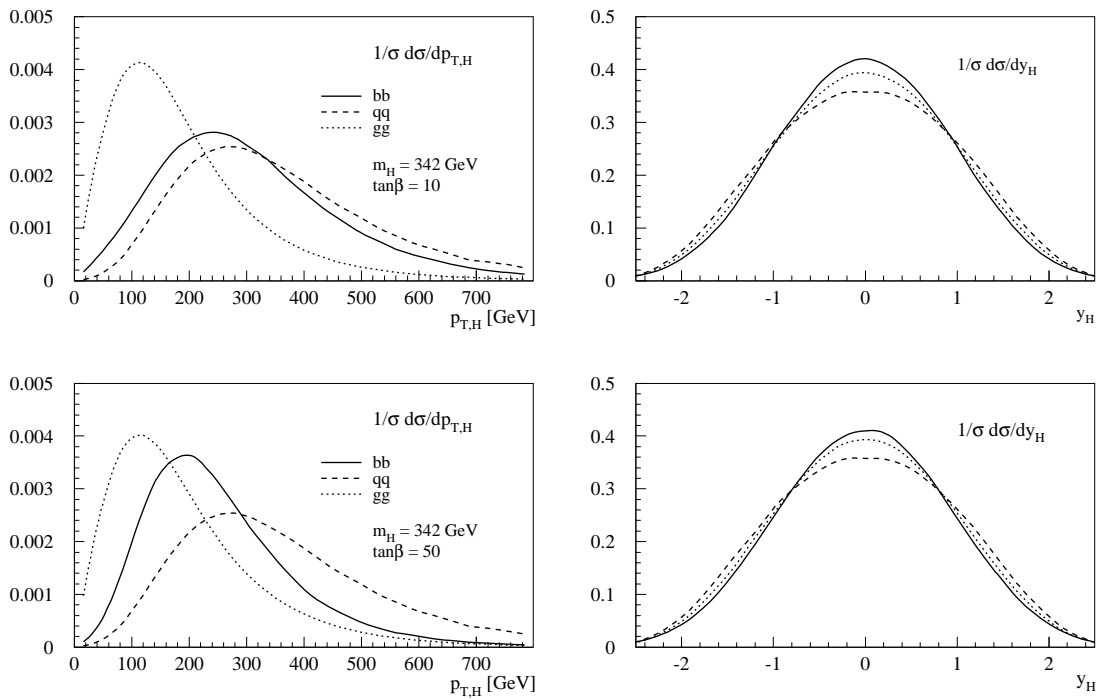}
 \caption[]{The rapidity and transverse momentum distributions of the
 final state charged Higgs boson for the three different production
 mechanisms. The curves for the bottom--induced process are labeled
 $bb$. The Drell--Yan process with only light-flavor incoming quarks
 is labeled $qq$, and the gluon fusion process through the one--loop
 amplitude is labeled $gg$. The bottom--induced production rate and
 the Drell--Yan production rates are computed at
 NLO.\label{fig:distri_all}}
\end{figure}

For large values of $\tan\beta$, the one--loop process $gg \to \hh$
cross section also increases like $\tan^4 \beta$. In the 2HDM this process
proceeds through a mixed top--bottom loop or though an $s$-channel
neutral Higgs boson. Even though our estimate of the gluon fusion rate
is less precise than for the other two channels, we use our LO 
estimate to compare the power of the different channels. The
theoretical uncertainty we would attach to this LO
cross section prediction is probably around a factor of two. However,
we believe that quoting the LO rate is conservative: we
expect the NLO corrections to be dominated by initial-state radiation
and therefore similar to the neutral Higgs pair production 
counterpart~\cite{neutral_pair}. As is seen in numerous processes, this
universal initial state radiation leads to large positive
corrections. We also note that this process is not evaluated with 
\msbar Yukawa couplings, but with their on-shell definition.
For loop--induced processes of this type, this
renormalization scheme leads to a well-behaved perturbative
series~\cite{spirix}.  The gluon fusion rate has a minimum around
$\tan\beta=7$, and in Fig.~\ref{fig:sig_all} we see that it passes the
(constant) Drell--Yan process at the LHC for $\tan\beta \gtrsim
40$. As stated before, the bottom--parton scattering rate is always
smaller than the loop--induced gluon fusion rate for production of
charged Higgs boson pairs at the LHC.\medskip

Na\"ively adding the rates for the three production mechanisms for
charged Higgs boson pair production is consistent, and it tells us which
process contributes what fraction of the events to the total cross
section. However, it will not be the entire story. Detector analyses
require acceptance cuts on the Higgs decay products, and these cuts
will also probe the distributions of the charged Higgs bosons. Therefore,
we have to compare the transverse momentum and rapidity
distributions of the charged Higgs bosons in the three different
production mechanisms. As before, we use the NLO results for the
bottom--parton scattering process and the Drell--Yan process. The
comparison is shown in Fig.~\ref{fig:distri_all}. As usual, we show
the results for two different values of $\tan\beta$, which distinguish
the $s$-channel ($\tan\beta=10$) and the $t$-channel ($\tan\beta=50$)
contributions. In contrast to bottom--parton scattering, we see
that the distributions for the Drell--Yan and gluon
fusion processes do not depend on $\tan\beta$. For Drell--Yan
$s$-channel photon and $Z$ exchange this reflects the fact that the
entire scattering matrix element is to a good approximation
independent of $\tan\beta$ (for large enough $m_H$). In gluon
fusion, the steep $x$ behavior of the gluon parton
distributions pushes the produced heavy particles close to threshold,
which means the two charged Higgs bosons appear with very little
transverse momentum. Again, this feature is independent of
$\tan\beta$. The bottom--parton scattering spectrum varies strongly
with the relative weight of the $s$-channel and the $t$-channel
diagrams. At threshold, $t$-channel top quark exchange ($S$--wave)
increases rapidly with $\beta$, much faster than $s$-channel gauge
boson exchange ($P$--wave), which shows a delayed increase, as 
$\beta^3$. The transverse momentum spectrum becomes softer the more
the $t$-channel contributes, \ie for larger values of $\tan\beta$. This
effect is particularly strong because the bottom parton densities are
derived from the gluon density and therefore are steep in $x$, pushing
the partonic center--of--mass energy to the smallest possible values.

\section{Supersymmetric Contribution to Bottom--Induced Process}
\label{sec:susy}

\begin{table}[t]
\begin{center}
\begin{tabular}{|c||cc|ccc|cc||cc||r|r|}
\hline
mSUGRA & $m_H$ & $\tan\beta$ & $m_0$ & $m_{1/2}$ & $A_0$ & $m_{\tilde{g}}$ & $\mu$ & $\sigma_{\rm LO}$[fb] & $\sigma_{\rm NLO}$[fb] & $\Delta_b$ & non--$\Delta_b$ \\
\hline
1a &  409  & 10 &  100 & 250 &   -100 &                       608 &  358 & $1.6\times 10^{-3}$  & $2.8\times 10^{-3}$  & -19\%  &   0.8\% \\
1b &  534  & 30 &  200 & 400 &      0 &                       938 &  516 & $4.1\times 10^{-3}$  & $5.5\times 10^{-3}$  & -40\%  &   2.7\% \\
2  & 1519  & 10 & 1450 & 300 &      0 &                       782 &  483 & $8.0\times 10^{-8}$  & $2.3\times 10^{-7}$  & -8.2\% &   3.1\% \\
3  &  595  & 10 &   90 & 400 &      0 &                       935 &  524 & $1.6\times 10^{-4}$  & $2.9\times 10^{-4}$  & -19\%  &   0.6\% \\
4  &  342  & 50 &  400 & 300 &      0 &                       733 &  396 & $5.3\times 10^{-1}$  & $6.0\times 10^{-1}$  & -54\%  &   3.9\% \\
5  &  698  &  5 &  150 & 300 &  -1000 &                       724 &  638 & $1.1\times 10^{-4}$  & $1.9\times 10^{-4}$  & -17\%  &   1.1\% \\
\hline      
GMSB & & & $\Lambda$ & $M_{\rm mes}$ & $N_{\rm mes}$ & & & & & & \\ 
\hline     
7  &  398  & 15 & $ 40 \times 10^3$ & $ 80 \times 10^3$ & 3 & 946 &  310 & $1.2\times 10^{-3}$  & $2.5\times 10^{-3}$  & -13\%  &   0.1\% \\    
8  &  543  & 15 & $100 \times 10^3$ & $200 \times 10^3$ & 1 & 835 &  423 & $2.0\times 10^{-4}$  & $4.0\times 10^{-4}$  & -13\%  &   0.2\% \\   
\hline
AMSB & & & $m_0$ & $m_{\tilde{G}} $ & & & & & & & \\ 
\hline         
9  & 1050  & 10 & 400  & $60 \times 10^3$ & &                1283 & 1022 & $2.5\times 10^{-6}$  & $5.4\times 10^{-6}$  & -25\%  &   1.6\% \\
\hline
\end{tabular}
\end{center}
\caption[]{Supersymmetric corrections to the production cross section
  $b\bb \to \hh$ from the resummed $\Delta_b$ corrections and 
  explicit remaining supersymmetric loop diagrams. The supersymmetric
  parameter points are chosen according to the benchmarks in
  Ref.~\cite{sps}.  All masses are given in units of GeV. The
  percentage changes are defined with respect to 2HDM rates at
  NLO.\label{tab:susy}}
\end{table}

Diagrams with a gluino instead of a gluon contribute to the NLO rate
for the process $pp \to \hh$. However, for massive gluinos only
virtual diagrams appear, in which the gluon and the quark lines are
replaced by gluinos and squarks. Usually, the SUSY contributions are
suppressed by the large supersymmetric partner masses. However, there
is a well-known exception to this rule, where the power suppression by
large sbottom masses can be compensated by the gluino mass and the
Higgsino mass parameter in the numerator. Sbottom--gluino corrections
to the external (massless) bottom legs can involve a left--right
mixing for the sbottom propagator. Because the off--diagonal matrix
element in the sbottom mass matrix has the form $m_b (A_b - \mu
\tan\beta)$ the left--right mixing is enhanced by a factor
$\tan\beta$. These potentially large corrections will
automatically become important for charged Higgs searches whenever we
rely on a large production rate with values $\tan\beta \gtrsim 30$ in
an MSSM framework. However, we also note that these corrections
introduce a huge new set of parameters into charged Higgs
searches, where it is not even clear if charged Higgs searches at
the LHC should be part of the search for supersymmetry or something more
general. We leave it to the reader to decide this issue, and present
the size of the additional supersymmetric corrections in this
section.\smallskip

Just like in mass renormalization, the supersymmetric external bubbles
can be resummed. It has been shown that this resummation consistently
takes into account the leading contributions in
$\tan\beta$~\cite{delta_b}. Note that in the large $\tan\beta$ limit
we approximate the off--diagonal matrix element as $-m_b \mu
\tan\beta$ and it is an unsolved problem how to treat the $m_b A_b$
term consistently. We will not resum this contribution. The difference
is in any case numerically small, because we are interested only in the
region of parameter space where $\tan\beta$ is large enough to also
ensure $\mu\tan\beta \gg A_b$. If we start from the fixed relation
between the bottom quark mass and the bottom Yukawa coupling (which we of
course break in our calculation with a vanishing bottom mass and a
running \msbar bottom Yukawa coupling) we can regard the resummed
bottom self-energy diagrams as a shift in the Yukawa coupling as
compared to the mass or its fixed order value:

\begin{alignat}{7}
\frac{m_b \tan\beta}{v} \; &\to \; \frac{m_b \tan\beta}{v} \;
                           \frac{1}{1+\Delta_b} \notag \\ \Delta_b &=
                           \; \frac{\sin(2 \theta_b)}{m_b} \;
                           \frac{\alpha_s}{4 \pi} \; C_F \;
                           m_{\tilde{g}} \; \frac{1}{i \pi^2} \;
                           \left[ B(0,m_{\tilde{b},2},m_{\tilde{g}})
                           -B(0,m_{\tilde{b},1},m_{\tilde{g}}) \right]
                           \phantom{haaallllooooooo} \notag \\ &= \;
                           \frac{\alpha_s}{2 \pi} \; C_F \;
                           m_{\tilde{g}} \; \left(- A_b + \mu \tan\beta
                           \right) \;
                           I(m_{\tilde{b},1},m_{\tilde{b},2},m_{\tilde{g}})
                           \notag \\[3mm] I(a,b,c) &= -
                           \frac{1}{(a^2-b^2)(b^2-c^2)(c^2-a^2)} \;
                           \left[ a^2b^2 \log \frac{a^2}{b^2} +b^2c^2
                           \log \frac{b^2}{c^2} +c^2a^2 \log
                           \frac{c^2}{a^2} \right] .
\end{alignat}
The function $B(p^2,m_1,m_2)$ is the usual scalar two--point function;
$C_F=4/3$ is the Casimir factor in the fundamental representation of
$SU(3)$. As mentioned above, there are similar additional terms
proportional to the strong coupling or to the top quark Yukawa
coupling, but the $\Delta_b$ correction is the leading contribution
for large $\tan\beta$. We should also note that the resummed
definition of these corrections is well defined for all $\Delta_b>-1$,
whereas the strictly fixed order series which yields $(1-4 \Delta_b)$
breaks down for $\Delta_b>1/4$.\smallskip

We show the numerical results for the supersymmetric corrections to
the total rate in Tab.~\ref{tab:susy}. To illustrate the effect of the
supersymmetric corrections we follow the SPS parameter points, which
are obviously not designed to advertise searches for charged Higgs
pairs in bottom--parton scattering. Apart from the fact that almost
all 2HDM production cross sections are very small, we indeed see that
the $\Delta_b$ corrections are by far dominant, and that the
remaining explicit diagrams contribute at a level well below the
total cross section theoretical uncertainty. The fact that all
$\Delta_b$ corrections are negative reflects the fact that all SPS
points are chosen with $\mu>0$, a preference triggered by indirect
constraints, such as measurements of the muon anomalous magnetic moment
$(g_\mu-2)$ or the rare decay $b \to s\gamma$.\medskip

Of course, all of the processes evaluated in section~\ref{sec:combo}
receive explicit supersymmetric corrections. For the Drell--Yan
process we know that their effects are small~\cite{prospino2}, because
there are no couplings enhanced by $\tan\beta$. For gluon fusion, 
the effects of squark loops can be very large if the squarks
are light. We do not include these additional diagrams in this paper,
because they require a careful analysis of the MSSM parameter
space. Instead, we refer the reader to dedicated studies in
Refs.~\cite{kniehl,obrian}. We should also mention that fairly light
charged Higgs bosons in an MSSM framework can be visible in cascade
decays of heavy squarks and gluinos~\cite{charged_susy}. Because the
production rates of squarks and gluinos are of the order of ${\cal
O}$(10 pb) at the LHC, even a small branching fraction to charged
Higgs bosons can be very promising.

\section{Conclusions}

To obtain a reliable prediction for the rate of charged Higgs boson pair
production at the LHC we have computed the process $b\bb \to \hh$ to
NLO. In previous studies~\cite{kniehl}, the bottom--parton scattering
process had been the source of large theoretical uncertainty. With our
improved understanding of bottom parton
densities~\cite{me_1,me_2,me_eduard}, we now have a recipe for
calculating processes involving initial-state bottom quarks. To test the
perturbative behavior of the LO prediction for $b\bb \to \hh$, 
we study NLO corrections to the total rate as well as to the
kinematical distributions of the heavy final state Higgs
bosons. Moreover, we use the NLO calculation to carefully verify the
validity of the bottom--parton approach.\smallskip

With this reliable prediction for the total cross section of the
bottom--induced production process at hand, we compare its rate to 
Drell--Yan pair production via an $s$-channel gauge boson. This
process is available at NLO through Prospino2~\cite{prospino2}. The
third production mechanism which we include in the comparison is the
gluon--fusion process with a one--loop amplitude including a mixed
top--bottom loop~\cite{charged_pair_gg}. Like the bottom--parton
scattering process its rate grows like $\tan\beta^4$ for large values
of $\tan\beta$. We find that for low values of $\tan\beta$ the
Drell--Yan process is by far dominant, and that for large values of
$\tan\beta$ the loop--induced process overwhelms the other two. 
Bottom--induced production will gain relative importance only if
supersymmetric loop contributions reduce the rate of the top--bottom
loop~\cite{kniehl,obrian}, or if the bottom--parton induced rate is
enhanced by universal supersymmetric corrections to the bottom Yukawa
coupling. Finally, we show how a combined sample of charged Higgs pair
events would look at the LHC, including the distinctly different
transverse momentum spectra of the three different production
processes.
\bigskip

\underline{Note added:} While we were in the process of finalizing
this work, another paper~\cite{other_pair} appeared in which the
authors present results which correspond to parts of our 
sections~\ref{sec:sigma} and \ref{sec:susy}.\bigskip

\section*{Acknowledgments}
 
We would like to thank Oscar \'Eboli for many enlightening discussions
and for his technical help. Moreover, TP would like to thank Oscar
\'Eboli and the University of São Paulo for their hospitality while
large fractions of the results presented in this paper were
calculated. Last, but not least we are very grateful to David
Rainwater and Oscar Éboli for their critical review of the manuscript.  
This work was supported in part by Funda\c{c}ão de Amparo \`a Pesquisa 
do Estado de São Paulo (FAPESP) through grant 04/01183-2 (AA).


\end{document}